\documentclass{raa}

\usepackage{graphicx,times}

\usepackage{natbib}

\bibpunct[, ]{(}{)}{;}{a}{}{,}

\def\beq{\begin{equation}}
\def\eeq{\end{equation}}
\def\bear{\begin{eqnarray}}
\def\ear{\end{eqnarray}}

\begin{document}

\title{Forming Different Planetary Systems}
 \volnopage{ {\bf 2012} Vol.\ {\bf 12} No. {\bf 8}, ~1081--1106}
   \setcounter{page}{1}

\author{Ji-Lin Zhou,  Ji-Wei Xie, Hui-Gen Liu, Hui Zhang, Yi-Sui Sun }
\institute{ Department of Astronomy and Key Laboratory of Modern Astronomy and
 Astrophysics in Ministry of Education,  Nanjing University,Nanjing 210093, China;
 zhoujl@nju.edu.cn \\
\vs \no
   {\small Received 2012.7.9; accepted 2012.7.17 }
}

\abstract{ With the increasing number of detected exoplanet samples, the statistical properties  of planetary systems have become much
clearer. In this review,  we summarize the major statistical results that have been revealed mainly by radial velocity and transiting
observations, and try to interpret them within the scope of the classical core-accretion scenario of planet formation, especially in
the formation of different orbital architectures for planetary systems around main sequence stars. Based on the different possible
formation routes for different planet systems, we tentatively classify them into three major catalogs: hot Jupiter systems, standard
systems and distant giant planet systems. The standard system can be further categorized into three sub-types  under different
circumstances: solar-like systems, hot Super-Earth systems, sub-giant planet systems.  We also review the planet detection and
formation in binary systems as well as planets in star clusters. }

   \authorrunning{Zhou et al.}
   \titlerunning{Forming Different Planetary Systems}
   \vspace{-3mm} \no{\sf INVITED REVIEWS}
   \maketitle

\section{Introduction}

Since the discovery of  a giant planet by radial velocity measurements in 51 Peg by \citet{Mayor95}, as well as the planets around 47
UMa and 70 Vir \citep{Butler96, Marcy96},  the new era of detecting exoplanets around  main sequence  stars was opened at the end of
last century. Now exoplanet detection has become one of the most rapidly developed area. With the development of measurement
techniques, the precision of radial velocity(RV) can be better than 0.5 $ms^{-1}$  with the HARPS spectrograph at La Silla Observatory
\citep{Mayor03}, making possible to detect  Earth-sized planets in close orbits around bright stars. The detection of transiting
signals when exoplanets pass in front of their host stars has become another powerful method in searching for planet candidates,
especially after the successful launch of CoRoT and Kepler . The unprecedented high precision of photometric observation ($\sim 10
$ppm) and long-duration continuous observation (up to years) achieved by space missions make transits  an ideal tool to detected
near-Earth-sized planets around solar type stars.

  To date, around 780 exoplanets  have been detected mainly by RV  measurements, with more than 100 multiple planet systems
\footnote{http://exoplanet.eu/,and hereafter for all RV based statistics in the paper.}. The first 16 months' observation of the
Kepler mission revealed more than 2321 transit candidates, with 48 candidates in the habitable zone of their host
stars\footnote{http://kepler.nasa.gov/,and hereafter in the paper} \citep{Borucki11,Batalha12,Fabrycky12}.

The study of planet formation can be traced  back to the 18th century, when  E. Swedenborg, I. Kant, and P.-S. Laplace developed the
nebular hypothesis for the formation of the solar system. At that time the solar system was the only sample of a planetary system. The
architecture of the solar system  implied that it was formed in a Keplerian disk of gas and dust (for reviews,  see
\citealt{Wetherill90, Lissauer95, Lin96}). With the discovery of more exoplanet systems, planet formation theory has developed
dramatically. For example, the discovery of hot Jupiters(HJs) stimulated the classical migration theory of planets embedded in the
proto-stellar disk \citep{Lin86, Lin96}, the moderate eccentricities of exoplanet orbits remind us the planet-planet
scattering(hereafter, PPS) theory \citep{Rasio96}, and the observation of some HJs in retrograde orbits extends the classical
Lidov-Kozai mechanism to eccentric cases \citep{Koz62,Lid62,Nao11b}. Through population synthesis, N-body  and hydrodynamical
simulations, the planet formation processes have been well revealed but are still far from fully understood.

 In this paper,  we focus on recent processes in the theory of detection and formation of solar type stars, either around single  stars (\S 2),
binary stars(\S 3), or in star clusters (\S 4).

\section{Planets around single stars}
\label{SG}

\subsection{Overview of Observations }

\subsubsection{Planet Occurrence Rate }

The occurrence rate of gas-giant exoplanets around solar-type stars has been relatively  well studied. \citet{Tabachnik02} fitted 72
planets  from different Doppler surveys,  and found a mass ($M$)-period ($P$) distribution with the form of a double power law, \beq
dN \propto  M^\alpha P^\beta d ln Md ln P, \label{dN} \eeq
 after accounting for selection effects. They obtained
$\alpha = -0.11 \pm  0.1$,  and $\beta =0.27 \pm 0.06$. Recently, \citet{Cumming08} analyzed eight years of precise RV measurements
from the Keck Planet Search, including a sample  of 585  chromospherically quiet stars  with spectrum types from  F to M.
 Such a power-law fit in equation (\ref{dN}) for planet masses $>0.3 M_J$ (Jupiter Mass) and periods $< 2000$ days was re-derived with $\alpha = -0.31 \pm  0.2$, and $\beta = 0.26 \pm 0.1$.
They  concluded $10.5\%$ of solar type stars have a planet with mass $>0.3 M_J$ and period $2-2000$ days,  with an extrapolation of
$17-20\%$ of stars having gas giant planets within 20 AU.

Based on the 8-year high precision RV survey  at the La Silla Observatory with the HARPS spectrograph, \citet{Mayor11} concluded  that
50\% of solar-type stars harbor at least one planet of any mass and with a period up to 100 days. About 14\% of solar-type stars have
a planetary companion more massive than $50M_\oplus$ in an orbit with a period shorter than 10 years. The mass distribution of
Super-Earths and Neptune-mass planets is strongly increasing between 30 and $15M_\oplus$, indicating small mass planets are more
frequent  around solar type stars.

\citet{Howard10} calculated  the occurrence rate of close planets (with $P< 50$ days), based on precise Doppler measurements of 166
Sun-like stars. They fitted the  measures as a power-law mass distribution, \beq df= 0.39 M^{-0.48} d\log M, \eeq
 indicating an increasing planet occurrence with decreasing
planet mass.  It also predicted that 23\% of stars harbor a close Earth-mass  planet, ranging from 0.5 to 2.0 $M_\oplus$ (Earth mass).

With  12 years of RV data with long-term instrumental precision better than $3 m s^{-1}$, the  Anglo-Australian Planet Search targets
254 stars, and estimates an occupance rate of  3.3\% of stars  hosting Jupiter analogs, and  no more than 37\% of stars hosting a
giant planet between 3 and 6 AU\citep{Rob11a}.
 Their results also reveal that the planet occurrence rate increases sharply with decreasing planetary mass.  In periods shorter than 50 days, they
  found that 3.0\% of stars host a giant ($m \sin i > 100M_\oplus$)
planet, and that 17.4\% of stars host a planet with ($m\sin i < 10M_\oplus$) \citep{Rob11b}.
Moreover,   the lack of  massive planets $(>8 M_{\rm Jup})$ beyond 4 AU was reported in \citet{Boisse12}, although with less than 20 RV detected candidates at the moment.
Such a distribution agrees with population synthesis \citep{Mordasini12} , where they showed that a decrease in frequency of massive giant planets at large distance ($\ge $ 5AU) is a solid prediction of the core-accretion theory.

   Transit observations from the Kepler  spacecraft give qualitatively similar results. \citet{Howard11b} checked the  distribution of planets in close orbits. For $P < 50$ days,  the distribution of planet radii ($R$) is given by a power law,
   \beq
   df= k (R/R_\oplus)^\alpha d \log R
   \eeq
   with $ k = 2.9^{+0.5}_{-0.4}$ and $\alpha = - 1.92 \pm  0.11$, and $R_\oplus$ is the Earth radius.
   They find the occurrence
of  $0.130 \pm 0.008$, $0.023 \pm 0.003$, and $0.013 \pm 0.002$ planets per star for planets with radii $2 -4$, $4-8 $, and $8 -32
R_\oplus$ respectively.

  The rapid increase of planet
occurrence with decreasing planet size indicates the presence of Super-Earth and Neptune size cases are quite common.  Although this
agrees with the prediction of the conventional core-accretion scenario,  it conflicts with the results predicted by the population
synthesis models that   a paucity of extrasolar planets with mass in the range $10-100 M_\oplus$ and semi-major
      axis less than 3 AU are expected, the so called 'planet desert' \citep{Ida04}.

\subsubsection{Stellar masses of planet hosting stars}

  Kepler results  also revealed   a correlation  between   the planet's occurrence   with
the effective temperature ($T_{\rm eff} $) of host stars \citep{Howard11b}.  The occurrence rate $f$ is inversely correlated with
$T_{\rm eff} $ for small planets with $R_p = 2-4 R_\oplus$, i.e.,
 \beq
  f= f_0 + k (T_{\rm eff} - 5100 K)/1000 K,
  \eeq
with  $f_0 = 0.165\pm  0.011$ and $k= - 0.081 \pm 0.011$. The
relation is valid over $T_{\rm eff} = 3600 -7100 K$,
 corresponding to stellar spectral typed from M ($\le 0.45M_\odot$, solar mass)to F($1.4M_\odot$).
This implies that stars with smaller masses tends to have small size planets.

  However, the occurrence of planets with radii larger than
$4 R_\oplus$ from Kepler does not appear to correlate with $T_{\rm eff}$ \citep{Howard11b}. This is in contrast with the observation
by RV measurements. In fact,   a much lower incidence of Jupiter-mass planets is found around M dwarfs
\citep{Butler06,Michael06,Johnson07},  and higher mass stars are more likely to host giant planets than lower-mass ones (e.g.,
\citealt{Lovis07,Johnson07,Johnson10,Borucki11}). The result is compatible  with the prediction  in the core accretion scenario for
planet formation \citep{Laughlin04,Ida05,Kennedy08}.

 \subsubsection{Metallicity Dependence}

 It has been well-established that more metal-rich stars have a higher probability
of harboring a giant planet than their lower metallicity counterparts
\citep{Gonzalez97,Santos01,Santos04,Fischer05,Udry07,Sozzetti09,Sousa11b}. The occurrence rate  increases dramatically with increasing
metallicity.  Based on the CORALIE and HARPS samples,  around 25\% of the stars with twice the metal content of our Sun are orbited by
a giant planet. This number decreases to $\sim  5\%$ for solar-metallicity cases \citep{Sousa11b,Mayor11}. Recently, \citet{Mortier12}
showed the  frequency of giant planets is $(f = 4.48^{+4.04}_{-1.38}\%)$ around stars with $[Fe/H] > - 0.7$,  as compared with $f\le
2.36\%$  around stars with $[Fe/H] \le  -0.7$. Curiously, no such correlation between planet host rate and stellar metallicity  is
observed for the lower-mass RV planets, and the stars hosting Neptune-mass planets  seem to have a rather flat metallicity
distribution \citep{Udry06,Sousa08,Sousa11b,Mayor11}.

By re-evaluating  the metallicity, \citet{Johnson09} find  that M dwarfs with planets appear to be systematically metal rich, a result
that is consistent with the metallicity distribution of FGK dwarfs with planets. \citet{Laughlin11} find that star hosting Kepler
exoplanet candidates  are preferentially metal-rich, including the low-mass stars that host candidates with small-radius , which
confirms the correlation between the metallicity of low-mass stars and the presence of low-mass and small-radius exoplanets.

\begin{figure}
\centering
\includegraphics[scale=0.7]{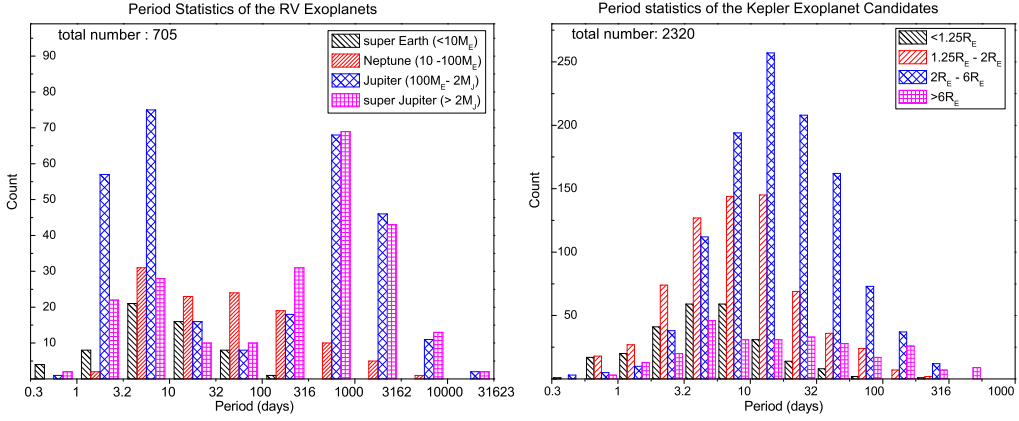}

\vspace{-3mm}
        \caption{\baselineskip 3mm
        Distribution of exoplanets detected by RV measure(left, http://exoplanet.eu/), and candidates by  the Kepler mission (right, http://kepler.nasa.gov/).  }
        \label{pot2}
\end{figure}

\subsubsection{Mass and Period Distributions}

Figure 1 shows the distribution of planetary orbital periods for different mass regimes.   705  planets detected by RV measurement and
2320 candidates revealed by the Kepler mission are included. Several features of the mass-period distribution have been well known and
widely discussed in the literatures. However, it seems that the distribution features from RV  detected exoplanets {\em are slightly
different} from those of Kepler candidates.

 \begin{itemize}
\item All kinds of RV planets show a  "pile-up" at orbital periods  2-7 days (e.g., \citealt{Gaudi05,Borucki11}), while
Kepler results show that Jupiter-size ($>6R_\oplus$) candidates are more or less flat up to orbits with $>100$ days;
  Neptune-size ($2M_\oplus<R_p<6R_\oplus$) and  super-Earth candidates ($1.25M_\oplus<R_p<2R_\oplus$)  peak at $10-20$days.
Both of them  are more abundanct relative to Jupiter-size candidates in the period range from one week to one month(Borucki et al.
2011). Extrapolating the distribution by considering the $(R_p/a)$ probability of a transiting exoplanet could extend   these  peaks
to a bit more distant orbits. Earth size or smaller candidates($<1.25M_\oplus$) show a peak of   $\sim 3$ days .

\item  RV planets show a paucity of massive planets ($M > 1M_J$) in close orbits \citep{Udry02,Udry03,Zucker02,zucker03}, and a  deficit of planets at intermediate orbital periods of
$10-100$ days \citep{Jones03,Udry03,Burkert07}. However, this is not obvious as Kepler results show at least several tens of
candidates with the radius lying in the $10-20 R_\oplus$ line  within 10-day orbits, and the distribution extending to 100 days is
rather flat. (Fig.1). The lack of all types of planets  with orbital periods $\sim 10-1000$ days observed by RV is clear, but from
Kepler results, the lack of planets with period $>100$ days is also shown, possibly due to the observational bias (Fig.1). RV
observations from the Anglo-Australian Planet Search  indicate that,  such a lack of giant planets ($M > 100M_\oplus$) with periods
between 10 and 100 days is indeed real. However, for planets in the mass range $10-100M_\oplus$, the results suggest that the deficit
of such planets may be a result of selection effects \citep{Rob10}.


\end{itemize}

\subsection{Hot Jupiter Systems}

The HJ class is referred  to a class of extrasolar planets and has mass close to or exceeding that of Jupiter ($M_p\ge 0.1M_{J}$, or
radius $\ge 8 R_\oplus$  for densities of $1.4 {\rm g/cm^{3}}$, a typical value of gas giants with small rocky cores),  with orbital
peroids $\le 10$ days  (or $a< 0.1$ AU) to their parent stars \citep{Cumming08,Howard11b,Wright12}.
 According to this definition, the RV exoplanets have 202 HJs, while Kepler candidates have 89 HJs.
  HJs are notable since they are easy to detecte either by RV or by transit measurements.  For example,  the
first exoplanet discovered around 51 Peg was such a close-in giant planet\citep{Mayor95}. Transiting HJs also give us information
about their radii, which is crucial for understanding their compositions(e.g., \citealt{Fortney03,Sara10} ).
 However, with the increase of RV precision and
 the number of detected exoplanets,  HJs are found to be in fact rare objects (e.g., \citealt{Cumming08,Wright12}).
 More interestingly,  some  HJs were observed in orbits that are retrograde with respect to the spin direction of their host stars(e.g., \citealt{Winn10}),
 indicating that their formation  process might have been quite different with that of our solar system.

\subsubsection{Occurrence rate and distributions}

 \citet{Marcy05a} analyzed 1330 stars from the Lick, Keck, and Anglo-Australian Planet Searches, and  the  rate of HJs among FGK dwarfs
surveyed by RV was estimated to be $1.2 \pm 0.1\%$. \citet{Mayor11} used the HARPS and CORALIE RV planet surveys  and found the
occurrence rate for planets with $M \sin i > 50M_\oplus$ and $P \le 11$ days is $0.89 \pm 0.36\%$. Recently, \citet{Wright12} used the
California Planet Survey from the Lick and Keck planet searches, and found the rate to be $1.2 \pm 0.38\%$. These numbers are more
than double the rate reported by \citet{Howard11b} for Kepler stars  ($0.5\pm 0.1\%$) and the rate of \citet{Gould06} from the
OGLE-III transit search. The difference might be, as pointed out by \citet{Wright12},  that transit surveys like OGLE and Kepler
(centered at galactic latitude $ b = +13.3^o$) probe a lower-metallicity population, on average, than RV surveys.

 Previous RV measurements show  that,  there is a sharp inner cutoff in the three day pileup of HJs. They appear to avoid the region inward of twice the
Roche radius \citep{Ford06}, where the Roche radius is the distance within which a planet would be tidally shredded. However, recent
RV detected exoplanets and Kepler candidates  indicate the presence of more than 200 exoplanet and candidates within 3-day orbits,
with the inner most orbital peroid being 0.24 days for system KOI-55, corresponding to a location close to its Roche radius (Fig.1).

Also, RV detected HJs appear to be less massive than more distant planets \citep{Patzold02,Zucker02}. For planets discovered with the
RV method, close planets have projected masses ($M \sin i$) less than twice Jupiter's mass. But numerous planets farther out have $M
\sin i > 2 M_J$ ( \citealt{Udry07}).

\subsubsection{ Spin-Orbit misalignment}
   One of the most fascinating features  for HJs is that,
some HJs  have orbits that are misaligned with respect to the spin of their host stars\citep{Winn10,Triaud10}. The sky-projected angle
between the stellar spin and the planet's orbital motion can be probed with the Rossiter-McLaughlin (RM) effect
\citep{Rossiter24,McLaughlin24}. To date,  the RM effect has now been measured for at least   47 transiting exoplanets (see
\citealt{Winn10}, Table 1 for a list of 28 planets, and \citet{Brown12}, Table 5  for a list of 19 additional  planets, and references
therein).  Only 7 (HAT-P-6b,HAT-P-7b, HATP-164b, WASP-8b,WASP-15b,WASP-17b,WASP-33) of the 47 samples have projected angles above
$90^o$, indicating a ratio of $\sim 15\%$ that are in retrograde motion. It is still not clear what type of stars could host HJs in
retrograde orbits. \citealt{Winn10} showed that the stars hosting HJs with retrograde orbits might have high effective temperatures
($>6250$ K). The underlying physics remans further study.

\subsubsection{Lack of  close companions}
 Few companion planets have been found in HJ systems within several AU \citep{Wright09,Hebrard10}.  To date, only six RV detected planetary system have  multiple planets with the
inner one being HJs  (HIP14810, ups And, HAT-P-13,HD187123, HD21707, HIP 11952 ). Compared to the total number of 89 RV detected  HJs,
the ratio is less than $7\%$.
  Interestingly, all these planetary companions are in  orbits with periods $> 140 $ days. This relative deficit also shows up in the transit samples, where
most attempts at detecting transit timing variations caused by close companions  have been unsuccessful
\citep{Holman05,Agol05,Rabus09,Csi10,Hrudkova10,Haghi11,Steffen12}). Kepler data also revealed the lack of a close companion in HJ
systems.
 \citet{Steffen12} present the results of a search for planetary companions orbiting near HJ
candidates  in the Kepler data.
Special emphasis is given to companions between the 2:1 interior and exterior mean-motion resonances(MMRS).
 A photometric transit search excludes the existence of
companions with sizes ranging from roughly $2/3$ to 5 times the size of Earth.


\subsection{Multiple Planet systems}

\begin{figure}
\centering
\includegraphics[scale=0.5]{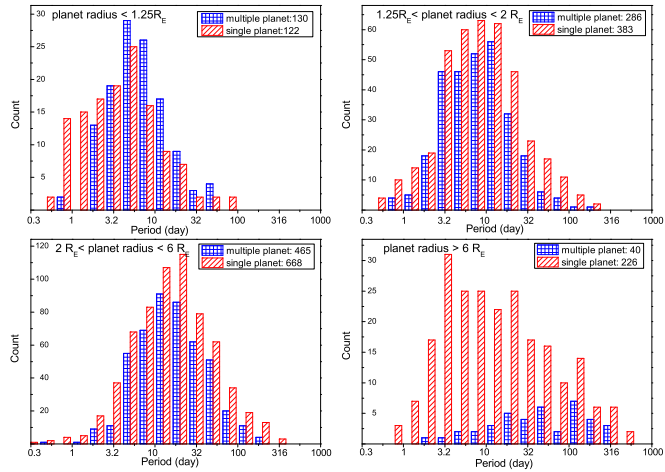}

\vspace{-3mm}
        \caption{\baselineskip 3.8mm
        The distribution of Kepler candidates in single and multiple planet systems for different  mass regimes. Data are from http://kepler.nasa.gov.   }
        \label{pot2}
\end{figure}

  With the increasing number of exoplanets being detected, the number of
multiple planet systems is also steadily increasing.  The first  16 months of Kepler data show that, among the 2321 candidates,  896
ones are in multiple planet systems, so that 20\% of the stars cataloged have multiple candidates\citep{Borucki11,Batalha12}.
Considering the present observation bias towards large mass planets, as well as the increasing occurrence rate of small mass planets,
we have a good reason to believe that multiple planets are very common and might occur at a much higher rate.
 The systems that have been revealed with the most numerous exoplanets are
  HD 10180 \citep[up to seven planets]{Lovis11} and
  Kepler 11 \citep[with six planets]{Lissauer11}.
  All of them are mainly composed of small mass  planets (Super Earth or Neptune mass).
   Several important signatures have been revealed by the Kepler mission:

\begin{itemize}
\item Multiple planets have on average smaller masses than single planet systems.
  Fig.2 shows  the paucity of giant planets at short orbital periods in multiple planet systems,
and the ratio of  giant planets (with radius $>6R_\oplus$) in single and multiple planet systems is roughly $5.7:1$, with orbital
periods of up to $\sim 500$ days' orbits.

\item Many planet pairs  are near  MMRs.
The presence of MMR  is a type of strong evidence for the migration history of the planet pairs (e.g., \citealt{LP02,Zhou05}).
\citealt{Wright11} summarized the data from RV detected planets, and found 20  planetary systems are apparently in MMRs, indicating
one-third of the well-characteried RV multiple planet systems have planet pairs in apparent MMRs.   \citealt{Fabrycky12}  found the
Kepler multiple transiting planet systems show some pile-up for planets pairs near lower order MMRs (especially 3:2 and 2:1 MMRs).

\item Multiple planet systems are nearly coplanar.
 Checking the Kepler multiple-transiting system indicates that these planets are typically coplanar to within a few degrees
 \citep{Batalha12}.
Also the comparison between the Kepler and RV surveys shows that
the mean inclination of multi-planet systems is less than $5^o$
\citep{Tremaine12}.
\citet{Figueira12} demonstrated that, in order to match the ratio of single planet systems to the 2-planet ones observed in HARPS and Kepler surveys, the distribution of mutual inclinations of multi-planet systems has to be of the order of $1^o$.

\end{itemize}

\subsection{Planet Formation Theory}

Now it is widely accepted that planets were formed in the
protoplanetary  disk during the early stage of star formation
 (e.g.,\citealt{Wetherill90,Lissauer95,Lin96,TF12}).  According to the
conventional core accretion model, planets are formed through the following processes (e.g., \citealt{Lis93,Arm07}):

 (1) grain condensation in the mid-plane of the gas disk, forming kilometer-sized planetesimals ($10^{18}-10^{22}$ g) on timescales on the order of $10^4$ yrs,
  from sticking collisions
of dust \citep{Weidenschilling93, weidenschilling97} , with gravitational fragmentation of a dense particle sub-disk near the midplane
of the protoplanetary disk \citep{Goldreich73,youdin02}. Further growth of planetesimals  can be helped by  procedures such as the
onset of streaming instability \citep{johansen07} or
 vortices  in turbulence \citep{cuzzi08},
or the sweeping of dust with the "snowball" model \citep{Xie10b,Ormel12,Win12}.

 (2) accretion of planetesimals into planetary
embryos ($10^{26}-10^{27}$ g, Mercury to Mars size) through a phase of ''runaway''  and ''oligarchic''  growth on a timescale of $\sim
10^4-10^5$ yrs \citep{Greenberg78, Wetherill89, Aarseth93,Kokubo96,Kokubo00,Rafikov03, Rafikov04}.

(3) gas accretion onto solid embryos with mass bigger than a
critical mass ($\sim  10 M_\oplus$) after a $\sim $Myrs long
quasi-equilibrium stage before gas depletion \citep{Mizuno80,
Bodenheimer86, Pollack96, Ikoma00}.

(4)  giant impacts between embryos, producing full-sized
$(10^{27}-10^{28}$ g) terrestrial planets in about $10^7-10^8$ yrs
\citep{cw98, la03, Kokubo06}. Thus the  presence of big solid
embryos and the lifetime  of the gas disk are crucial for the
presence of giant plants, while the presence of enough heavy
element  determines the mass of solid embryos and terrestrial
planets.


According to the above scenario, the correlation between stars that host giant planets and stellar metallicity can be understood.
 By cosmological assumption, a high stellar metallicity implies a protoplanety disk with more heavy elements,  thus a metal-rich protoplanetary disk
  enable the rapid formation of  Earth-mass embryos necessary to form the cores of giant planets before the gaseous disk is dissipated.
   That correlation might also
indicate a lower limit on the amount of solid material necessary to form giant planets. \citet{Jarrett12} estimated a lower limit of
the critical abundance for planet formation of $[Fe/H]_{\rm crit} \sim -1.5 +\log({\rm r}/1{\rm AU})$,  where $r $ is the distance to
the star. Another key point may be the correlation between  metallicity and the  lifetime of the gas disk. There is observational
evidence that the lifetime of circumstellar disks is short at lower metallicity, likely due to the great susceptibility to
photoevaporation\citep{yasui09}.

Although the above procedures for single planet formation are relatively clear, there are some bottleneck questions (see previous
listed reviews).  Next, we focus on  the formation of orbital architectures for different planet systems.

\subsubsection{Formation of Hot Jupiter systems} \label{SHJ}

Due to the high temperature that might hinder the accretion of gas in forming giant planets, the HJs were assumed to be formed in
distance orbits rather than formed {\em in situ}.
 There are mainly three theories that were proposed to
explain the formation of HJ systems with the observed
configurations.

{\em Disk migration model}. The earliest model for the formation of HJ systems  is the planet migration theory embed in protostellar
disks \citep{Lin86, LBP96}. Giant planets formed in distant orbits, then migrated inward under the planet-disk interactions and
angular momentum exchanges\citep{GT80,Lin86}.  The so called type II migration will be stalled  at the inner disk edge truncated by
the stellar magnetic field.
 The  maximum distance of disk  truncation is estimated to be $\sim 9$
 stellar radii \citep{Kon91}. Considering  the radius of the protostar is generally 2-3 times
  larger than their counterpart in the main sequence, the inner disk truncation would occur at $\sim 0.1$AU.
This might naturally explain the pileup of orbits with periods of $3-10$ days' for HJs. However, as type II migration is effective
only in the plane of the disk, and disk's tidal forces try to dampen the inclination of planets \citep{GT80}, this procedure can not
explain the formation of HJs in orbits with high inclination, as well as the lack of planetary companions in close orbits. Recently,
\citet{lai11} proposed that stellar-disk interaction may gradually shift the stellar spin axis away from the disk plane, on a time
scale up to Gyrs.

{\em Planetary scattering model}. Another  mechanism that might  account for the formation of HJ systems is the PPS model.
Close-encounters among planets can excite their orbital eccentricities ($e$). In the extreme case that $e$ is near unity,  the orbital
periastron will be small enough so that star-planet tidal interactions might be effective and  circularize the orbits to become HJs
\citep{Rasio96, ford01, pt01, FR08}. The planetary scattering model can reproduce the observed eccentricity distribution of moderately
eccentric $(e\sim 0.1-0.3)$, non-HJ extra-solar planets \citep{zhou07, Cha08, jt08}.
 However, 
 the required high eccentricity and the long timescale required for tidal damping be effective
 might be not easy to achieve unless some secular effects (e.g. the Lidov-Kozai mechanism) are excited (e.g., \citealt{Nag08}).

{\em Secular models}.  The third class of models relate to the Lidov-Kozai effect \citep{Lid62,Koz62} in the presence of a third body.
To account for the high inclination of HJs,  \citet{Wu03} proposed that a companion star which is a third body in a high inclination
orbit can induce Kozai oscillations on the planet's evolution, gradually exciting the planet's orbit to an eccentricity near unity  so
that it can reach a proximity close to the central star, until tidal dissipation circularizes the orbit into
a HJ. 
 \citet{FT07} found such resulting HJs should be
double-peaked with orbital inclinations of $\sim 40^o$ and $140^o$. Such an idea has been extended to brown dwarf companion by
\citet{Nao11a}.

 However, because the population studies have established that only $10\%$ of HJs can be explained by Kozai migration due to binary companions
\citep{wu07, FT07},  but studies show that most of the HJ systems do not have stellar or substellar companions. Whether this mechanism
can account for the formation of most  HJs is not known. Another question is that,   in the stellar companion case ($m_c$, a star or a
brown dwarf), the  orbital angular momentum (AM) of $m_c$ dominates that of the system and determines an invariant plane, thus the
$z-$component of AM (perpendicular to the invariant plane) of the planet ($m_p$)  is conserved when $m_c$ is in a distant orbit. Thus
$m_p$ can be in an apparent retrograde orbit relative to the spin axis of the main star only when $m_c$ has a relatively large
inclination with respect to the equator of the main star\citep{wu07, FT07}, and this retrograde motion is {\em not} with respect to
the invariant plane determined by the total AM.

To avoid relying on the effects of stars or brown dwarf companions,  and also to find the occurrence of retrograde motion relative to
the invariant plane, one resorts to the conditions under which the Lidov-Kozai mechanism works for planet mass companions ($m_c$).
\citet{Nao11b} study the mechanism with a general three-body model. Denote $a,a_c$  as the semimajor axes of inner planet ($m_p$) and
companion, respectively, with $e_c$ being the eccentricity of $m_c$; they find that as long as $(a/a_c)e_c/(1-e_c^2)$ is not
negligible, the octuple-level of the three-body Hamiltonian would be effective, so that the z-component of $m_p$ in AM  is no longer
conserved, allowing the occurrence of retrograde motion relative to the invariant plane.
 Thus , to make  a retrograde HJ,   a companion in a close and  eccentric orbit is required,
  but the mass of the companion is not important.

However,  newly-born planets are assumed to be in near circular and coplanar orbits.  To generate the required eccentricity,
\citealt{Nag08,Nag11} introduced planet scattering into the above pictures. Starting from a relatively compact system  ($\sim 3.6R_H$,
Hill's radius) with three Jupiter-mass planets, the planets scatter one another on a timescale of $\sim 10^3$ years. They found $\sim
30\%$ of the simulations can result in a planet with eccentricity high enough, that  Kozai excitations  from  outer planets can become
effective, so that it can be either in a close orbit with non-negligible  eccentricity, or in a highly inclined (even a retrograde
orbit) with relatively small eccentricity over a timescale of $10^9$ years. However, it is unclear whether  the initial condition of a
compact and highly unstable planetary system can exist, as required by this theory \citep{Matsumura10}.  Also the scatted planets can
be observed to test the theory.

Another route  to generate eccentricities other than through violent PPS is the diffusive chaos arising from a multiple planet system
after it forms. The generation of eccentricity in a multiple planet system is a slow, random walk diffusion in the velocity dispersion
space, and the timescale increases with the logarithm of the initial orbital separations \citep{zhou07}. Recently, \citealt{Wu11}
proposed that {\em secular chaos}  may be excited in an orderly space system, and it  may  lead to natural excitation of the
eccentricity and inclination of the inner system, resulting in observed HJ systems. They inferred that such a theory can also explain
the eccentricities and inclinations for distant giant planets. However, to what extent such a mechanism could be effective within the
age of planetary systems remains for further study.


To summarize, the Lidov-Kozai mechanism seems to be the most promising mechanism for the formation of HJs.  Provided that  initial
eccentricities of the planet's companion can reach high enough value, inter planetary Kozai oscillations can bring the inner planets
into HJ orbits with sufficiently high inclinations.


\subsubsection{Formation of multiple planet architectures and a system of classification}


What should a 'standard' planet system be like? Before answering this question, let us first check the possible outcome of a planet
system after the formation of individuals  by the procedure listed at the beginning of section 2.4.

According to the core accretion scenario, by depleting all the heavy elements in a nearby region (called the feeding zone, roughly 10
Hill radii), an embryo without any migration will be  stalled from growing, which is a case called {\em  an isolation
mass}\citep{Ida04}. In a disk with metallicity $f_d$ times the minimum mass solar nebula(MNSN) \citep{Hayashi81}, the isolation mass
can be estimated as (\citealt{Ida04}, Eq.19) \beq
 M_{\rm iso} \approx 0.16 \eta_{\rm ice}^{3/2}f_d^{3/2}(\frac{a}{\rm 1AU})^{3/4}(\frac{M_*}{M_\odot})^{-1/2}M_\oplus,
\eeq where $\eta_{\rm ice}$ is the enhancement factor, with a value of $1$ and $\approx 4.2$ respectively inside and outside the snow
line ( location with temperature  170K beyond which water is in the form of ice, $\sim 2.7$ AU in the solar system). The time required
for the core to accrete nearby materials and become isolated is on the order of  (\citealt{Ida04}. Eq.18) \beq
 \tau\approx 1.2\times 10^5 \eta_{\rm ice}^{-1}f_d^{-1}f_g^{-2/5}(\frac{a}{1\rm AU})^{27/10}
 (\frac{M_{\rm iso}}{M_\oplus})^{1/3}(\frac{M_*}{M_\odot})^{-1/6}~{\rm yr},
 \eeq
 where $f_g$ is the enhancement factor of gas disk over MMSN.
 So for a typical disk with 2 times the MNSN ($f_d=f_g=2)$,  isolation embryos  inside the snow line are small ($ < 1 M_\oplus$), and they can not develop. Embryos
beyond the snow line can grow to $\sim 10M_\oplus$ so that they can accrete gas to form gas giants. However, the growth time of
embryos with mass $10M_\oplus$ in distant orbits ($>20AU$) is long  ($\sim 10Myr$ at 10AU and $\sim 70Myr$ at 20 AU).
 Within  a  disk with a moderate lifetime of $\sim 3$ Myr for
classical T-Tauri stars \citep{Haisch01} ,    embryos in distant orbits do not have enough time to accrete gas, thus they will stall
their growth at the mass of a sub-giant mass, like Uranus and Neptune in the solar system.

 As  the gas disk is depleted,
the induced secular resonance sweeps through the inner region of the planetary systems, causing further  mergers of cores
\citep{Nag03}.
 Terrestrial planet formed after the
gas disk was depleted  at $\sim 200$Myrs\citep{Cha01} .  After depletion of the gas disk , a debris disk with leftover cores
interacted with giant plants, causing small scale migration, such as in the Nice model\citep{Gomes05,Morbi05,Tsiga05}. Thus,  assuming
no giant migrations occurred, the solar system is the basic "standard" multiple planet system. As all planetary embryos were formed in
near mid-plane of the gas disk, without  perturbations in the vertical direction, such a  standard planet system is nearly coplanar,
like many multiple planet systems observed by the Kepler mission.

However, several procedures make the above picture more complicated. One of the most difficult task is to understand
the migration of embryos or planets embedded in the gas disk before depletion.
 For  a sub-Earth protoplanet , the exchanges of angular momentum between it
and the nearby  gas disk will cause a net momentum lose on it, which results in a so-called type I migration over a timescale on the
order of $< 0.1Myr$ \citep{Gold79, Ward86, Ward97, Tanaka02}. If the protoplanet can avoid such disastrous  inward migration, and
successfully grow massive enough to accrete gas and become a gas giant, the viscous evolution of the disk may cause the giant planet
 to undergo a type II migration, with a timescale of Myrs\citep{Lin86}. Recent studies infer that, under more realistic conditions, the
migration speeds of both types  can be reduced or even with their direction-being reversed, leading to an even rarer outcome
\citealt{Kley12}).


The evidence for planet migration is the observed systems in MMR. Since 2:1 MMR has the widest resonance width, especially for
planetesimals in nearly circular  orbits \citep{MD99}, many planet pair are expected to show 2:1 MMRs if they had a history of
convergent migration (e.g.  \citealt{Rivera10,Zhou10, WS12,GH12}). However, Kepler planets give many planets conditions near but not
in MMR. This can be understood by the phenomenon that later stage planetesimal and planet interactions may cause further migrations
but with smaller extensions, causing  strict commensurability to be lost \citep{CP07}. Giant planets in MMR might be strong enough and
survive under such perturbations, like the GJ 876 system\citep{LP02,Rivera10,GH12}. Hydrodynamical simulations show that different
disk geometries might lead the planet pair to either  convergent migration (thus  possibly the trap of different MMRs), or sometimes
to divergent migrations\citep{ZZ10a,ZZ10b}. However planet pairs may not necessarily lead to MMRs configurations for some dynamical
configurations \citep{BM12}, e.g. the resonant repulsion of planet pairs is discussed by \citet{LW12}.

 The orbital configurations of multiple planet systems incorporating planetary migration have been studied  extensively by population syntheses (e.g., \citealt{Ida08,Ida10},\citealp{Mordasini09a,Mordasini09b}) and N-body simulations(e.g.,\citealt{Thom08},  \citealt{Liu11}). \citet{Thom08}
 found that for giant planet formation,  two timescales are crucial: the lifetime of the gas disk $\tau_{\rm disk}$ and the time to form the first gas giant $\tau_{\rm giant}$.
 In cases with $\tau_{\rm giant} > \tau_{\rm disk}$,  the gas is removed before any gas giant
has a chance to form, leaving behind systems consisting solely of rocky-icy bodies. In cases
with $\tau_{\rm giant} < \tau_{\rm disk}$,
such systems generally produced a number of gas
giants that migrated inward a considerable distance.
 \citet{Liu11}  also showed that $\tau_{\rm disk}$ is crucial for forming planet systems, as large $\tau_{disk}$ tends to form  more giant planets in close and nearly circular orbits, while small $\tau_{\rm disk}$ favors forming planets  with small masses in distant and eccentric orbits.

According to the above theories as well as the  currently available observations, the  planet systems around solar type stars can be
roughly classified into the following categories. A detailed classification will be presented later( Zhou et al. , in preparation).

 \begin{itemize}

\item {\em Class I: Hot Jupiter systems}. These might be formed through some secular mechanisms such as  Lidov-Kozai cycling , as discussed previously.  Typical example: 51 Peg b.

\item {\em Class II: Standard systems }.  They are formed either through processes similar to our solar system, or by undergoing some large scale migrations, as mentioned perviously.
According to scenarios, they undergo migration, due to the deficit of heavy elements in the gas disk, or due to the short lifetime of
disk, They can further be classified as,

\begin{itemize}
\item {\em Subclass II-1: Solar-like  systems.}  These have planetary configurations similar to the solar system: terrestrial planets in the inner part, 2-3 gas giant planets in middle orbits, and Neptune-size sub-giants in outer orbits, due to insufficient gas accretion.  Typical example: Mu Ara, ups And, and HD125612 systems.

\item {\em Subclass II-2: Hot super-Earth systems.}  With the migration of giant planets,  the sweeping of inward  MMRs  or secular resonances
will trap the isolated masses ($0.1-1M_\oplus$), and excite their eccentricities , causing  further mergers, which result in the
formation of hot super Earth, like GJ 876d  \citep{Zhou05, Raymond06, Raymond08}.   Other formation scenarios, see a review
\citep{Hag11}. Typical example: GJ 876, and  Kepler 9 systems.

\item {\em Subclass II-3:  Sub-giant planet systems}. Due to the low disk mass or low metallicity,
 planet embryos around some stars (especially M dwarf) might not grow massive enough to accrete sufficient  gas to become a gas giant,
  thus  planets in these systems are generally sub-giants, like most of the  systems discovered in \citet{Mayor11}.  Typical example:  the Kepler 11 system.

\end{itemize}
\item {\em Class III.  Distant giant systems}.
 Through direct imaging, a type of system was detected with many massive companions (up to  several times the mass of of Jupiter) in distant orbits,
such as Fomalhaut b \citep{Kala08} ,  the HR8799 system\citep{Maro08}, and beta Pic b \citep{Lag09}. Interestingly, all these stars
have short ages $(\sim 100-300 $ Myrs). Whether the planets were formed in situ through gravitational instability\citep{Boss97}, or
formed through outward migration or scattering, is still not clear. Typical examples: Fomalhaut, HR 8799, and beta Pic systems.
\end{itemize}

\section{Planets in Binary Star Systems}
\label{PBSS}
\subsection{Overview of Observation }
Planets in binaries are of particular interest as most stars are believed to be born not alone but in a group, e.g.,  binaries and
multiple stellars systems.
 Currently, the  multiplicity rate of solar like stars is $\sim 44-46\%$, including $\sim 34-38\%$ for only binaries \citep{DM91, Rag10}.
 Different resulting values of the multiplicity rate of planet-bearing stars (compared to all the planet hosts) were  found to be  $23\%$ \citep{Rag06}
  and $\sim17\%$ \citep{MN09}, and most recently $\sim12\%$ \citep{Roe12}. The decreasing multiplicity rate is mainly because of the quickly increasing
  number of transiting planets discovered in recent years. For example, Kepler has discovered more than 60 planets since 2010, however, followup
  multiplicity studies on such planet hosts are usually postponed or even considered impracticable.  In any case, the multiplicity rate of a planet host is
  significantly less than the multiplicity rate of stars. This may be because of selection biases in planet-detection against binary systems and/or
  because of impacts of binarity on planet formation and evolution \citep{Egg11}.

Depending on the orbital configuration, planets in binaries are usually divided into two categories  \citep{Hag10,HagB10}, S type for
planets orbiting around one of the stellar binary components, i.e.,the circumprimary case, and P type for those orbiting around both
the stellar binary components, i.e., the circumbinary case.  Currently, most of them are S type, and only a few are found in P type,
including NN Ser \citep{Beu10}, HW Vir \citep{Lee09}, DP Leo \citep{Qia10}, HU Aqr \citep{Qia10, Hin12}, UZ For \citep{Dai10, Pot11},
Kepler-16 (AB)b, Kepler-34 (AB)b, and Kepler-35 (AB)b \citep{Doy11, Wel12}. In the following, we will focus more on the former, and a
binary system, hereafter, refers to S type unless explicitly noted otherwise.

According to the most recent summary of observations \citep{Roe12}, there are 57 S type planet-bearing binary systems \footnote{In
fact, 10 of them are triple stellar systems,  but with the third star being very far away and thus exerting less effects on the
binaries with planets}, which, as a subsample of extra-solar planetary systems, may provide some significant statistics. Here we
summarize several points worth noting.
\begin{enumerate}
\item \emph{Binary separation (or orbital semimajor axis, $a_{B}$).} Most S type systems have a $a_{B}$ larger than 100 AU.  However, there seems to be a pileup at $a_{B}\sim20$ AU with 4 systems: $\gamma$ Cephei \citep{Hat03}, Gl 86 \citep{Queloz00b}, HD 41004 \citep{Zucker04}, and HD196885 \citep{Cor08, Cha11}. Planets are slightly less frequent in binaries with  $a_{B}$ between 35 and 100 AU \citep{Egg11}. No planet has been found in binaries with $a_{B}<10$ AU (excluding P type).

\item \emph{Planetary mass}. Planets in wide binaries ($a_{B}>100$ AU) has a mass range ($0.01-10 M_{J}$)  that is close to those in single star systems but much more extended than those ($0.1-10 M_{J}$) in close binaries ($a_{B}<100$ AU) \citep{Roe12}.

\item \emph{Planetary multiplicity}. Planets in close binaries ($a_{B}>100$ AU) are all singleton, while those in wide binaries are diverse (Fig.3 of \citet{Roe12}). The occurrence rate of multiple planets in wide binaries is close to that in single star systems \citep{DB07}.

\item \emph{Planetary orbit}. Most extremely eccentric planets are found in wider binaries (e.g., $e=0.935$ for HD 80606 b and $e=0.925$ for HD 20782 b).  The distribution od planetary eccentricity in binaries also seems to be different compared to those in single star systems \citep{Kai12}. Planetary orbital periods are slightly smaller in close binaries as compared to  those in wide binaries and single star systems \citep{DB07}.

\end{enumerate}

How are these planets formed with double suns? Are they behaving in a similar way as our solar system or other single star systems? In
the following, we review some important effects on planet formation and evolution in a binary system as compared to those in a single
star system, which may provide some clues to answer these questions.

\subsection{Binary Effects on a Protoplanetary Disk}

\subsubsection{Disk Truncation}
Planets are considered to be born in a protoplanetary disk. Such a disk, in the solar system, could be extended to the location of the
Kuiper belt, e.g., 30-50 AU from the Sun. But in a binary system, the disk could be severely truncated by the companion star. For the
S type case, the typical radial size of a truncated disk is about $20-40\%$ of the binary's separation, depending on the mass ratio
and orbital eccentricity of the binary. For the P type case, the binary truncates the circumbinary disk by opening a gap in the inner
region. The typical radial size of the gap is about 2-5 times the binary's separation distance. Various empirical formulas for
estimating the boundary of the truncated disk are given by \citet{AL94, HW99, Pic05}\footnote{The boundaries given by \citet{HW99} and
\citet{Pic05} are actually the boundaries of stable orbits of a test particle.} .  The size range of the truncated disk puts the first
strict constraint on planet formation, determining where planets are allowed to reside and how much material is available for their
formation. The reason why no S-type planet has been found in binaries with $a_{B}<10$ AU could be that the truncated protoplanetary
disk was too small to have enough material for formation of a giant planet \citep{Jan07}.

\subsubsection{Disk Distortion}
After the violent truncation process, the left-over, truncated disk, is still subject to strong perturbations from the companion star,
and thus it is not as dynamically quiet as disks around single stars. First, a binary in an eccentric orbit can also cause the disk to
be  eccentric \citep{Paa08, KN08, MK12}. Second, if the binary orbital plane is misaligned with respect to the disk plane, then binary
perturbations can cause the disk to become warped, twisted or even disrupted \citep{Lar96, FN10}. Third, the eccentric, warped disk is
precessing. All the above effects cause planet formation in binary systems to be more complicated than that in single star systems.

\subsubsection{Disk Lifetime}
Estimating the lifetime of the protoplanetary disk is crucial as it provides a strong constraint on the timescale of planet formation.
Observations of disks around single stars show that the typical disk life time is in the range 1-10 Myr \citep{Haisch01}. Although
disks around wide binaries show a similar lifetime, those in close binaries ($a_{B}<40$ AU) show evidence of shorter lifetime, i.e.,
$\sim 0.1-1$ Myr \citep{Cie09}. Such results are not unexpected as disks in close binaries are truncated to a much smaller size and
thus have much smaller timescales of viscous evolution. In any case, such a short disk lifetime requires that planets in close
binaries (such as $\gamma$ Cephei) should form quickly, probably on a timescale less than 1 Myr.

\subsection{Binary Effects on Planet Formation}
We consider planet formation based on the core accretion scenario \citep{Lis93, Cha04} \footnote{Gravitational instability is another
candidate scenario for planet formation in binaries (see \citet{May10} for a review)}, starting from planetesimals (usually having a
radius on the scale of kilometers) embedded in a protoplanetary disk. This is the standard way that people consider planet formation
in single star systems, though planetesimal formation itself is still unclear \citep{BW08, CY10}. Nevertheless, some observational
indications imply that the first stages of planet formation, i.e., dust settling and growing to planetesimals, could proceed in
binaries as well as in single star systems \citep{Pas08}.

\subsubsection{Growing Planetesimals}
One straightforward way for growing planetesimals is via mutual collisions and mergers, as long as the collisional velocity $V_{col}$
is low enough. For a protoplanetary disk around a single star system, if the disk turbulence is weak, e.g., in a dead zone, growth by
mutual collisions could be efficient, and it is thought that planetesimals have undergone a runaway and oligarchic phase of growth to
become planetary embryos or protoplanets \citep{Kokubo96, Kokubo98}. However, the situation becomes less clear in binary systems. On
one hand, the outcome of planetesimal-planetesimal collision is highly sensitive to $V_{col}$ \citep{BA99, SL09}. On the other hand,
perturbations from a close binary companion can excite planetesimal orbits and increase their mutual impact velocities, $V_{col}$, to
values that might exceed their escape velocities or even the critical velocities for the onset of eroding collisions \citep{Hep78,
Whi98}. This is a thorny problem for those binaries with separation of only $\sim 20$ AU, such as $\gamma$ Cephei and HD 196885.
Recently, many studies have been performed to address this issue.

An earlier investigation by \citet{MS00} found that the combination of binary perturbations and local gas damping could force a strong
orbital alignment between planetesimal orbits, which significantly reduced $V_{col}$  despite relatively high planetesimal
eccentricities. This mechanism was thought to solve the problem of planetesimal growth until \citet{The06} found the orbital alignment
is size-dependent. Planetesimals of different sizes align their orbits to different orientations, thus $V_{col}$ values between
different sized planetesimals are still high enough to inhibit planetesimal growth (\citet{The08, The09} for S-type, and \citet{Mes12}
for P type). Moreover, the situation would become much more complicated (probably unfavorable) for planetesimal growth if the
eccentricity, inclination and precession of the gas disk are also considered \citep{Cie07, Paa08, Mar09, Bea10, Xie11, Fra11, Bat11,
Zha12}. Nevertheless, the problem could be somewhat simplified if the effects of a dissipating gas disk are taken into account
\citep{XZ08} and/or a smaller inclination ($i_{B}< 10^{\circ}$) between the binary orbit and the plane of the protoplanetary disk is
considered \citep{XZ09}. Optimistically, planetesimals may undergo a delayed runaway growth mode (called Type II runaway) towards
planets \citep{Kor01}. In any case, however, it seems that planetesimal-planetesimal collision is not an efficient way for growing
planetesimals in close binary systems.

An alternative way of growing planetesimals could be via accretion of dust that they pass through in the disk. Both analytical studies
and simulations \citep{Xie10b, PL10, Win12} have shown this could be promising to solve the problem of growing planetesimals not only
in binaries but also in single star systems (e.g., the well known ``meter-barrier'' puzzle). For an efficient dust accretion to occur,
one needs, first, a source of dust, which could be either from the primordial protoplanetary disk or from fragmentation of
planetesimal-planetesimal collisions, and second, weak disk turbulence to maintain a high volume density of dust \citep{Joh08}.

\subsubsection{Formation of Terrestrial and Gaseous Planets}
Once planetesimals grew to 100-1000 km in radius (usually called planetary embryos or protoplanets), they are no longer as fragile as
before. Their own gravity is strong enough to prevent them from fragmenting by mutual collisions. In such a case,  most collisions
lead to mergers and thus growth of planetesimals. Hence, one way to speed up growth is by increasing $V_{col}$, which is readily
available in  a binary star system. For close binaries, such as $\alpha$ Centauri AB, simulations \citep{Bar02, Qui04, QL06, Qui07,
HR07,Gue08} have shown that Earth-like planets could be formed in 10-100 Myr.

If a protoplanet reaches several Earth masses, the critical mass for triggering a runway gas accretion, before the gas disk is
depleted, then it could accrete the surrounding gas to become a gaseous planet. Generally, planets would stop gas accretion after they
have cleared all the surrounding gas and opened a gap. However, because of the binary perturbation, gas could be pushed inward to
refill the gap and finally accreted by the planet \citep{Kle01}, leading to a higher gas accretion rate and more massive gaseous
planets. Such an effect could partially explain  one of the observed facts: gaseous planets in close binaries are slightly more
massive than those in single star systems.

 \subsection{Binary Effects on Planetary Dynamical  Evolution}
\subsubsection{With a Gas Disk}
Due to the complication of the problem itself, the studies of this aspect mainly rely on numerical simulations. \citet{KN08}
considered the evolution of a low-mass planet (30 Earth masses) embedded in a gas disk of the $\gamma$ Cephei system (S type). They
found that the planet would rapid migrate inward and accrete a large fraction of the disk's gas to become a gas giant planet, which is
similar to the observed planet. For the circumbinary case, i.e., P type, simulations \citep{PN07, PN08a, PN08b} showed the results
were sensitive to planet mass.  Low mass planets (tens of Earth masses) would successively migrate inward to the inner edge of the gas
disk and subsequently merge, scatter, and/or lock into a MMR. A high mass planet ($>$ Jupiter mass) would enter a 4:1 resonance with
the binary, which pumped up the eccentricity of the planet and probably led to instability.  The model favoring the low mass planet
from the simulation is consistent with the recent observation: the masses of the three confirmed circumbinary planets
(Kepler-16,-34,-35) are all $\le$ Saturn's mass.

\subsubsection{Without a Gas Disk}
As the lifetime of the disk, typically $\le$ 10 Myr,  is only less than 1\% of that of a planet (typically on the order of Gyr), the
subsequent gas free phase could dominate the evolution of planets after they have formed. In fact, several mechanisms are found to
play an important role in shaping the final structure of planetary systems in binaries.
\begin{itemize}
\item \emph{Planet-planet scattering}.
Multiple planets could form in a protoplanetary disk, and because of damping from the gas disk , planets could maintain their near
circular orbits and thus avoid close encounters. Once the gas disk dissipated, planet-planet interaction would excite the
eccentricities of planets, leading to close encounters and finally instability of the systems; e.g., merger and/or ejection. Such a
mechanism (usually called PPS) is thought to explain the eccentricity distribution of observed giant planets \citep{FR08, Nag08,
Cha08}.  In a binary system, PPS would be more violent because of the additional perturbation from the binary stars \citep{Mal07a}. In
S type binaries (especially those with close separations and highly inclined and/or eccentric orbits), simulations (\citet{Mar05}, Xie
et al. in prep.)  have shown that PPS often causes the system to finally have only a single planet, and the remaining planet is
usually the most massive one. Such results may  explain one observed fact:  planets in close binaries are single and massive. In P
type binaries, PPS again favors a single planet. In addition, it predicts a positive correlation between the planet's orbital
semimajor axis and eccentricity (Gong et al. in prep.), which currently fit well to the three confirmed circumbinary planets
(Kepler-16,-34,-35). More P type planets detected in the future will further test this correlation.

\item \emph{Lidov-Kozai Effect}.
In an S type binary, if a planet is on a highly inclined orbit\footnote{This could be either primordial or induced by planet-planet
scattering.}, then it could undergo the Lidov-Kozai effect \citep{Koz62, Lid62}. One of the most striking features of the effect is
that the planet's eccentricity could be pumped to a very high value and oscillate with its inclination out of phase.  Recently, it has
also been found that the planet could flip its orbit back and forth when its eccentricity approaches unity \citep{LN11, Nao11a} if the
binary orbit is eccentric, hence exhibiting the so called eccentric Lidov-Kozai effect.  One application of this effect is that it
could produce a HJ; when the planet oscillates to very high eccentricity, with a very small periastron, tides from the central star
kick in and dampen its orbits to form a close planet \citep{Wu03, FT07}. Recently, there have been examples of such candidates showing
evidence that they are on the way to being HJs via the Lidov-Kozai effect \citep{Soc12, DJ12}. In addition, as the planet could flip
during the Lidov-Kozai evolution, there are significant chances to form an HJ in a retrograde orbit \citep{Nao11a}, which has been
observed in some extrasolar systems \citep{Triaud10}.  Nevertheless, depending on specific conditions, e.g, if general relativistic
effects and/or perturbation by another planet is relevant, the Lidov-Kozai effect can be suppressed \citep{TR08, SR09}.

\end{itemize}

\subsection{Non-Primordial Scenario}
There is another possibility that a currently observed planet-bearing binary was not the original one when the planet was born, namely
the non-primordial scenario. Various mechanisms can lead to such a result, and we briefly summarize these two kinds as follows.

\begin{itemize}
\item \emph{Encounters with other stars and/or planets.}
A binary star system has a larger collisional cross section than a single star and thus a larger chance to  have a close encounter
with other stars, during which they could have their planets lost and/or exchanged \citep{Pfa05, MB12}. In the end, the binaries
probably dramatically changed their orbits, and the surviving planets were probably excited to highly eccentric and/or inclined
orbits\citep{Mal07b, Spurzem09, Mal11}.  In addition, free floating planets(FFPs) could be recaptured by flyby binary stellar systems
\citep{PK12}.

\item \emph{Steller Evolution.}
If one of the binary component star evolves away from the main sequence, it could induce instabilities in the planetary system in the
binary. Planets could bounce back and be forced between the space around the two component binary stars \citep{Kratter12}. If a close
binary star evolve to some phase to have mass transfer, the mass lost from the donor star could form a circumbinary disk, which could
potentially harbor new planets \citep{Per10}.

\end{itemize}

\section{Planet in star Clusters}
\label{SC}

Almost all the planets found now are around field stars. However the normal theory of star formation predicts that most field stars
are formed from a molecular cloud, having the same initial mass function(IMF) as stars($<3M_{\odot}$) in an open cluster indicating
that these field stars initially formed in clusters, e.g. our solar system's, initial birth environment was reviewed by
\cite{Adams10}. According to the chemical composition of our solar system, our Sun may have formed in an environment with thousands of
stars, i.e. a star cluster or association. Thus scientists are very interested in planet detection in clusters which would be more
effective than that around field stars due to many more objects existing in the same size of a telescope's field of view.

To survey planets around stars in a cluster, we have some advantages in obtaining more effective and credible results. Some
correlations between planets occurrences as well as their properties and characteristics of their host stars are not very clear due to
the bias of measurements for these field stars, such as age, mass, [Fe/H] etc. Large differences among these field stars, especially
the type of environment in the early stage, is a problem for surveying the correlations. However, in one cluster, most of its members
have homogeneous physical parameters, i.e. age and [Fe/H]. The comparative study of planets in clusters will provide more valid,
credible correlations.

Unfortunately, except for some FFPs, few planets are found to be bounded around members of either globular clusters(GCs) or open
clusters(OCs). The following sections will introduce the observational results and theoretical works in both GCs and OCs respectively.

%
%

\subsection{Planets in Globular Clusters}

Because of the fruitful observation results of GCs and the huge number of stars in GCs, especially main sequence stars(MSSs), people
naturally expect to find planets around these MSSs in GCs. As these stars are, on average about 50 times denser than field stars near
the Sun, GCs have advantages for planet searching. For example, in the two brightest GCs: $\omega$ Cen and 47 Tuc, there are more than
60000 MSSs, approximately half of the total number of Kepler targets. However, the extreme star density near the center of GCs ($10^5$
stars within a few arc min), requires an extremely high precision of photometry. Until now, it has been hard to individually
distinguish two nearby stars in the core of GCs. The stars in the outer region of GCs are more widely separated from each other,
therefore they are more suitable for planet searches.

The first planet system in GC was found in the nearest GC: M4, PSR B1620-26 b \citep{Backer93}, a $2.5 M_{\rm J}$ planet around a
binary radio pulsar composed with a $1.35 M_\odot$ pulsar and a $0.6 M_\odot$ white dwarf. However, if we focus on  sun-like stars in
GCs, no planets have been confirmed until now.

To search for bounded planets around MSSs, some efforts have been made by several groups. As the brightest GCs in the sky, 47 Tucanae
and $\omega$ Centauri are good targets for planet searching by transiting. Using HST to find planets in the core of 47 Tucanae,
\cite{Gilliland00} provided a null result. In the outer halo, the same result was obtained by \cite{Weldrake05}. Furthermore,
\cite{Weldrake07} found no bounded planets by transiting in both of the two clusters, under the precision of $P < 7 $day, $1.3-1.6
R_{\rm J}$. The most recent works to find planets in the nearby globular cluster NGC 6397 is contributed by \cite{Nascimbeni12},
butstill no highly-significant planetary candidates have been detected for early-M type cluster members.


Do the null results in GCs indicate the low occurrence of planets? For some dense star environments, the stability of planets is
crucial. Although planets at $1 $AU in the core of 47 Tuc can only survive around $10^8 $yrs in such a violent dynamical
environment\citep{Davies01}, planets at $10$AU in the uncrowded halo of GCs can be preserved for several Gyrs \citep{Bonnell01}.
Therefore HJs with periods around a few days can survive much longer in the halo. If HJs formed near these cluster members, they would
have a chance to be detected in GCs \citep{Fregeau06}.

The null results are mainly attributed to the low metallicity of these GCs. \cite{Fischer05} surveyed planet systems not far from the
Sun, and pointed out that the occurrence of gas giants depends on the metallicity of their host stars. The most recent work by
\cite{Mortier12} found a frequency of HJs $< 1\%$ around metal-poor stars, while the frequency of gas giants is $<2.36\%$ around stars
with $[Fe/H]<-0.7$. Both 47 Tuc and $\omega$ Cen have a low [Fe/H](respectively -0.78 and $<-1$, from data collected by
\citet{Harris96}). Hence, these two GCs contains few HJs. Higher frequencies of giant planet are expected in GCs with higher [Fe/H].

Additionally, different properties of a circumstellar gas disk, especially its structure, might influence the final architecture of
planet systems, e.g. if the gas disk in GCs is depleted much faster due to Extreme-Ultraviolet(EUV) and Far-Ultraviolet(FUV)
evaporation from nearby massive stars \citep{Matsuyama03}, formation of gas giants may be unlikely, as well as the formation of a hot
planet. The different structure of a gas disk might not force the planet to migrate inward enough to form hot planets, and naturally
they are hard to detect using transit.

In these old GCs, mass segregation is obvious due to energy equipartition, i.e. massive objects concentrate in the center of the
cluster while small objects are easily ejected outside. Energy equipartition results in FFPs, which might be ejected to become unbound
by some mechanisms \citep{Parker12, Veras12} and have a lower-mass than stars. It is hard for FFPs to stay in old GCs.

\subsection{Planets in Open Clusters and Associations}


None of the planets around solar-like stars are found in GCs,because of the reasons mentioned before. OCs and associations which still
contain lots of MSSs, are also useful for planet searching. The main dissimilarities between OCs and GCs are:
\begin{enumerate}
\item \emph{Cluster ages}.
OCs and associations are much younger than GCs, and have a much larger [Fe/H], probably leading to more planets being formed around
the cluster members.

\item \emph{The dynamical environments}.
The dynamical environment in OCs and associations is still less violent than that in GCs because of lower star density, which can
preserve the two-body systems more easily than in GCs.

\item \emph{Binary fraction}.
The much larger fraction of binary systems in OCs than GCs is a good way to
understand the formation of planet systems in binary stars.

\end{enumerate}
Additionally, many more OCs and associations($\sim1200$) are observed than GCs ( $\sim 160$ ) in our galaxy.   Due to these
dissimilarities, a higher probability of planet detection is expected.


As for the different properties of OCs, surveyed planets in OCs have their own values. Some OCs are only a few Myr old, e.g.
NGC6611\citep{Bonatto06} and NGC 2244 \citep{Bonatto09}. Their ages are comparable with the timescale of planet formation. Surveying
planets and circumstellar disks in these very young clusters will provide valuable samples to check and enhance the current theories
of planet formation, particularly the influences via different environments in clusters during the early stages of planet formation.

\subsubsection{Bounded Planets and Debris Disks}

Many groups have made efforts to search for planets by transits in OCs: e.g. \cite{Bruntt03} in NGC 6791, \cite{Bramich05} in NGC
7789, \cite{Rosvick06} in NGC 7086, \cite{Mochejska06} in NGC 2158, etc. Only few candidates were found but none were confirmed. The
most significant progresses were made in 2007. In NGC 2423, a gas giant with a minimum mass of $10.6 M_{\rm J}$ around a $2.4 M_\odot$
red giant was found by \cite{Lovis07} using RV  measurement. Another planet was soon found soon afterward by RV around the giant star
$\epsilon ~$Tauri (\cite{Sato07}) in the Hyades, the nearest OC, with a minimum mass of $\sim7.6M_{\rm J}$ and a period of $\sim 595 $
days. Using transit, some smaller candidates have also been found without RV confirmation, e.g. a single transit of a candidate
$\sim1.81 M_{\rm J}$ in NGC 7789 found by \cite{Bramich05}, which may indicate another exoplanet with a long period. Most recent work
by Quinn et al. (2012) claims that they found two HJs by RV: Pr0201b and Pr0211b in Praesepe, these planet are the first known HJs in
OCs. Parameters describing these planets are listed in Table 1 as well as properties of their host cluster.
\begin{table}
\begin{center}
\caption{Parameters of the planets and their host stars in OCs, as well as their host cluster. Data are from
\cite{Lovis07,Sato07,Harris96,Quinn12}\label{tbl-1}}.
\begin{tabular}{ccccccccc}
\hline\hline\\
$M_{\rm p}\sin i$ &  Period & semi-major & ecc  & $M_{\rm star}$ & host cluster & age& [Fe/H]& dist\\
$(M_{\rm J}) $ & (day) & axis (AU) & &$M_\odot$ & &(Gyr) & &(pc)\\
\hline\\

$10.6$ &$714.3$ &$2.1$ &$0.21$ &$2.4$ &$NGC 2423$ &$0.74$ &$0.14$ &$766$\\
$7.6$ &$594.9$ &$1.93$ &$0.15$ &$2.7$ &$Hyades$ &$0.6$ &$0.19$ &$47$\\
\hline
\end{tabular}
\end{center}
\end{table}

Compared with the null results in GCs, the encuraging results of planet searching in OCs confirm the formation and survival ability of
planets in cluster environment, especially observations of the circumstellar disk in young OCs, which is related to the occurrence of
planet formation.

\cite{Haisch01} showed the fraction of disks in OCs decayed with their ages. Some recent results verify this correlation: $30-35\%$ of
T-Tauri stars have a disk in the $\sigma$ Ori cluster with ages $\sim 3 $ Myr\citep{Hernandez07}.  Using the Chandra X-ray
Observatory, \cite{Wang11} found a K-excess disk frequency of $3.8\pm0.7\%$ in the $5\sim10 r$My old cluster: Trumpler 15.

Although the disk structures around cluster members are not well known, the large fraction of gas disk in very young OCs makes the
 formation of planets possible, especially for gas giants. The two confirmed planets found were not HJs, but another two planets found most
recently are HJs. However, lack of more samples is a big problem in making a credible conclusion and surveying the statistical
characteristics of planet formation and evolution in OCs.

\subsubsection{Free-Floating Planets}

Ages, metallicity and star density are the main dissimilarities between OCs and GCs. The formation of planets in OCs is thought to be
common, but few planets bound around stars  have been observed. However a population of FFPs has been found in OCs. In 2000,
\cite{Lucas00} found a population of FFPs in Orion. \cite{Bihain09} also found three additional FFPs with $4-6M_{\rm J}$ in the $\sim
3$ Myr old OC: $\sigma $~Orions. A huge number (nearly twice the number around MSSs) of unbound planets have been found in the
direction of the Galactic Bulge \citep{Sumi11}.

These planets have multiple origins. They may also form around some cluster members, but were ejected out of the original systems and
cruise into clusters \citep{Sumi11}. Because of their young ages, energy equipartition in OCs is less effective than that in GCs. The
dissolution timescale for objects to escape from a cluster is $t_{\rm dis}\sim 2{\rm Myr}\times\frac{N}{\ln(0.4N)}\times\frac{R_{\rm
G}}{\rm kpc}$ \citep{Baumgardt03}. For a typical OC, with $N=1000$ stars at distance $R_{\rm G}=1$kpc, $t_{\rm dis}\sim 0.1$Gyr and
therefore FFPs can still stay in their host clusters for most young OCs. It is hard to find the original host stars of these FFPs, but
surveying them is still useful for evaluating the frequency of planet formation in OCs and GCs.

\subsection{Planetary Systems in Clusters: theoretic works}

The planet occurrent rate including formation rate and stability related with the cluster environment is very important for predicting
the rate of further observations. From their respective dynamics, the large distinctions between OCs and GCs generally predict more
planets in OCs and HJs in halo of GCs.

Dynamical works focus on the stability and orbital architecture of planetary systems in a cluster. Considering a fly-by event, the
previous works show the stability of planet systems depends on the bounded energy of planetary systems, fly-by parameters as well as
the star density of the environment, which decides the occurrent rate of a fly-by event ($t_{\rm enc}\propto 1/n$,
\citealt{Binney87}). \cite{Spurzem09} used a strict N-body simulation as well as a Monte Carlo method to survey the dynamical
characteristics, especially the effective cross section of planetary systems with different orbital elements in a cluster's gravity
field. Adding substructure of a young OC by \cite{Parker12}, the fraction of liberated planet depends on the initial semi-major axis
and virial parameter. The planet systems in binary systems were also  been surveyed by \cite{Mal07a,Mal07b,Malmberg09}. They
considered encounters between a binary system and a single star. After obvious changes of the inclination, a fraction of planets will
suffer the Kozai effect after encounters and consequently show instabilities.

The stability and orbital architecture of multi-planet systems in clusters still need to be surveyed in further works, because
planet-planet interactions play an important role in deciding the final configuration of a planet system after fly-bys. The dynamical
evolution in clusters is much more complex than in a single fly-by. In some very open clusters, the tidal effect can also disrupt
planet systems in the outer region. The effect of interstellar gas in very young OCs is still uncertain. The fine structure of the
circumstellar disks still needs to be investigated during the formation of planet systems.




Planet formation in star clusters must have a strong dependence on the physical and dynamical environments of their host stars. The
environments in clusters are very different from that around field stars, or binary pairs, e.g. the different properties of the
circumstellar disk, dynamical instabilities in different stages during planet formation, as well as the stability of a planetary
system after the planets are formed. The protoplanet gas disk plays a very important role in the formation of gas giant planets. A
comparison between the timescale of gas disk dispersion and that of gas giant formation is a crucial clue to judge the formation rate
of giant planets. On the other hand, the observation of circumstellar disks and giant planets (including FFPs) in some very young OCs,
can also give a limit on the rate that a planetary gas disk is preserved, which is related to the planet formation rate in a cluster
environment. The distinctions in the different environments for OCs and small bounded planet samples in OCs have limited our knowledge
about the formation of planets in clusters.

\section{Conclusions}

With the increasing data of observed exoplanets, the study of orbital architectures for multiple planet systems becomes timely. Unlike
the relatively mature theory for formation of a single planet (except for some bottleneck problems), the properties of planet's
architecture is relatively far from clear.  Dynamical factors, such as interactions among planets, tidal interactions with the host
star and a protostellar disk, or in some cases perturbations from a third companion (a star or brown dwarf), etc, tend to sculpt the
orbital evolution and sculpt  the final architectures of the planet systems.

According to our present knowledge, we tentatively classify the planet systems around single stars into three major catalogs: HJ
systems, standard systems and distanct giant planet systems. The standard systems can be further categorized into three sub-types
under different circumstances:  solar-like systems, hot super-Earth systems and sub-giant planet systems. The classification is based
on the major process that occurred in their history. It may help to predict unseen planets, as well as to understand the possible
composition of planets, since through the history of their evolution, we can judge whether large orbital mixing has occurred.

Due to the presence of a third companion, planet formation in a binary environment has raised some more challenging problems,
especially for the stage of planetesimal formation. Anyway, the observed exoplanets around binary stars, especially the circumbinary
exoplanets like Kepler 34b and 35b,  indicate that planet formation is a robust procedure around solar type stars.

Planets in clusters will provide a useful clue for understanding the formation of planets in a cluster environment. Although only very
limited observational results have been obtained, theories can still predict some properties of exoplanets in clusters. Planet samples
in some young OCs might be especially interesting for revealing the difference between planet formation around field stars and members
of clusters.

\normalem
\begin{acknowledgements}
We would like to thank  Drs. G.Marcy,  S.Udry, N.Haghighipour,M.Nagasawa,  Y.Q. Wu, A.V.Tutukov and R.Wittenmyer, I.Boisse for their
helpful suggestions. The work is supported by the National Natural Science Foundations of China (Nos.10833001,10925313,11078001 and
11003010), Fundamental Research Funds for the Central Universities(No.1112020102), the Research Fund for the Doctoral Program of
Higher Education of China(No.20090091110002 and 20090091120025).
\end{acknowledgements}

\end{document}